\documentclass[twocolumn,aps,prb,showpacs,superscriptaddress,longbibliography]{revtex4-1}

\usepackage{graphicx}
\usepackage{dcolumn}
\usepackage{bm}
\usepackage{color}
\usepackage{siunitx}
\usepackage{placeins}

\begin{document}

\title{Driven Dipolariton Transistors in Y-shaped Channels }

\author{Patrick Serafin}
\affiliation{Physics Department, New York City College of Technology, The City University of New York, Brooklyn, New York 11201, USA}

\author{Tim Byrnes}
\affiliation{New York University Shanghai, 1555 Century Avenue, Pudong, Shanghai 200122, China}

\author{German V. Kolmakov}
\affiliation{Physics Department, New York City College of Technology, The City University of New York, Brooklyn, New York 11201, USA}

\date{\today}

\begin{abstract}
Exciton-dipolaritons are investigated as a platform for realizing
working elements of a polaritronic transistor.  Exciton-dipolaritons are 
three-way superposition of cavity photons, direct and indirect excitons in a bilayer
semiconducting system embedded in an optical microcavity.  
Using the forced diffusion equation for dipolaritons, we study the room-temperature dynamics of
dipolaritons in a transition-metal dichalcogenide (TMD) heterogeneous bilayer.  
Specifically, we considered a MoSe$_2$-WS$_2$  heterostructure, where a Y-shaped channel guiding the dipolariton
propagation is produced. We demonstrate that polaritronic signals can be redistributed in the channels by applying a driving voltage in an optimal direction.  Our findings open a route towards the design of an efficient room-temperature dipolariton-based optical transistor.

\end{abstract}

\maketitle

\section{Introduction}
 Semiconductor technology has allowed for the innovation of devices such as light emitting diodes \cite{tsintzos:5, jayaprakash:7}, lasers \cite{Bajoni:3,khalifa:6}, and polarization rectifiers \cite{sedov:8}. The emerging field of polaritronics has opened the route towards the use of polaritons in many innovative structures \cite{Sanvitto:4}. Research into exciton-polariton dynamics has allowed for the observance of their rich physical phenomena, such as Bose-Einstein condensate superfluidity \cite{Alberto:5}, quantum vortices \cite{lagoudakis:16}, lasing capabilities \cite{Christopoulos:10}, and novel many-body physics \cite{byrnes:91,byrnes:92,byrnes:93}. In addition, exciton-polariton transport has been investigated in a variety of novel materials \cite{Masumoto:15,ma:14,ardizzone:9} and has sparked interest in the study of their manipulation and trapping within semiconductor microcavities \cite{kim:02}. The potential for use of exciton-polaritons  in a wide array of applications such as quantum information \cite{Byrnes:22}, polariton lasers \cite{fraser:05}, tunneling diodes \cite{Nguyen:14}, and optical transistors \cite{zasedatelev:11} has led to study and review of their properties and phenomenology \cite{byrnes:17}. Atomtronics, the atomic counterpart of polaritronics, has led to various quantum technology applications \cite{Amico:10} such as atomtronic batteries \cite{Caliga:11} and circuits \cite{pepino:12}. In this report we narrow our attention to the study of polaritronics, specifically, we investigate exciton-polariton dynamics. 

 The emergence of transition metal dichalcogenides (TMDs)  provide unique opportunities to realize many proposed applications \cite{Gao:85,Berman:10}  of exciton and polariton physics to room-temperature scales. Unique properties of TMDs include  large exciton binding energy $\sim 1$ eV \cite{He:2} and large vacuum Rabi splitting energy $\sim 100$ meV  \cite{Vinod:6}. 
Tungsten based-TMDs heterostructures provide near degenerate interlayer and intralayer excitionic states, which enables one to dynamically tune these states via the application of an electric field normal to the layers or the gate voltage \cite{Ceballos:7}.
In particular, it has been found that TMD heterostructures possess band alignments that allow for spatial separation between the electrons and holes in the TMD layers upon lasing \cite{Ceballos:7}.  

In this paper,  we study propagation of polaritons in MoSe$_2$-WS$_2$  heterogeneous bilayer structures embedded in an optical microcavity. Dipolaritons are quasiparticles with a three-way superposition of
cavity photons, intralayer (direct) and interlayer (indirect) excitons as seen in Fig.1  \cite{Cristofolini:12,Byrnes:14}. Additionally, the band diagram for MoSe$_2$-WS$_2$ provides for a clear visual representation of the formation of dipolaritons present in our system \cite{Bilayer:51}.
The advantage of dipolaritons lie in the possibility to directly apply an electric force due to the driving 
electric voltage applied to one of the layers \cite{Su:08}.  The latter makes dipolaritons a promising candidates for polaritronic, optoelectronic, and photonic applications \cite{Byrnes:14}.
In our approach, the heterostructure possesses a Y-shaped channel guiding the dipolariton propagation.
Y-shaped channels were recently proposed to guide exciton polariton propagation in gallium-arsenide (GaAs) based microcavities at liquid helium temperatures \cite{Berman:14}. In the present work, we numerically investigate an electrically controlled optical switch  based on dipolaritons in MoSe$_2$-WS$_2$ heterostructures in an optical cavity, which are able to operate at room temperatures \cite{calman:11,Choi:23}. In particular we show the optimal system parameters that enable the efficient re-routing of dipolaritons in the Y-shaped channel of the TMD heterostructure.

 \begin{figure}[h]
\begin{center}
\includegraphics [height=6cm]{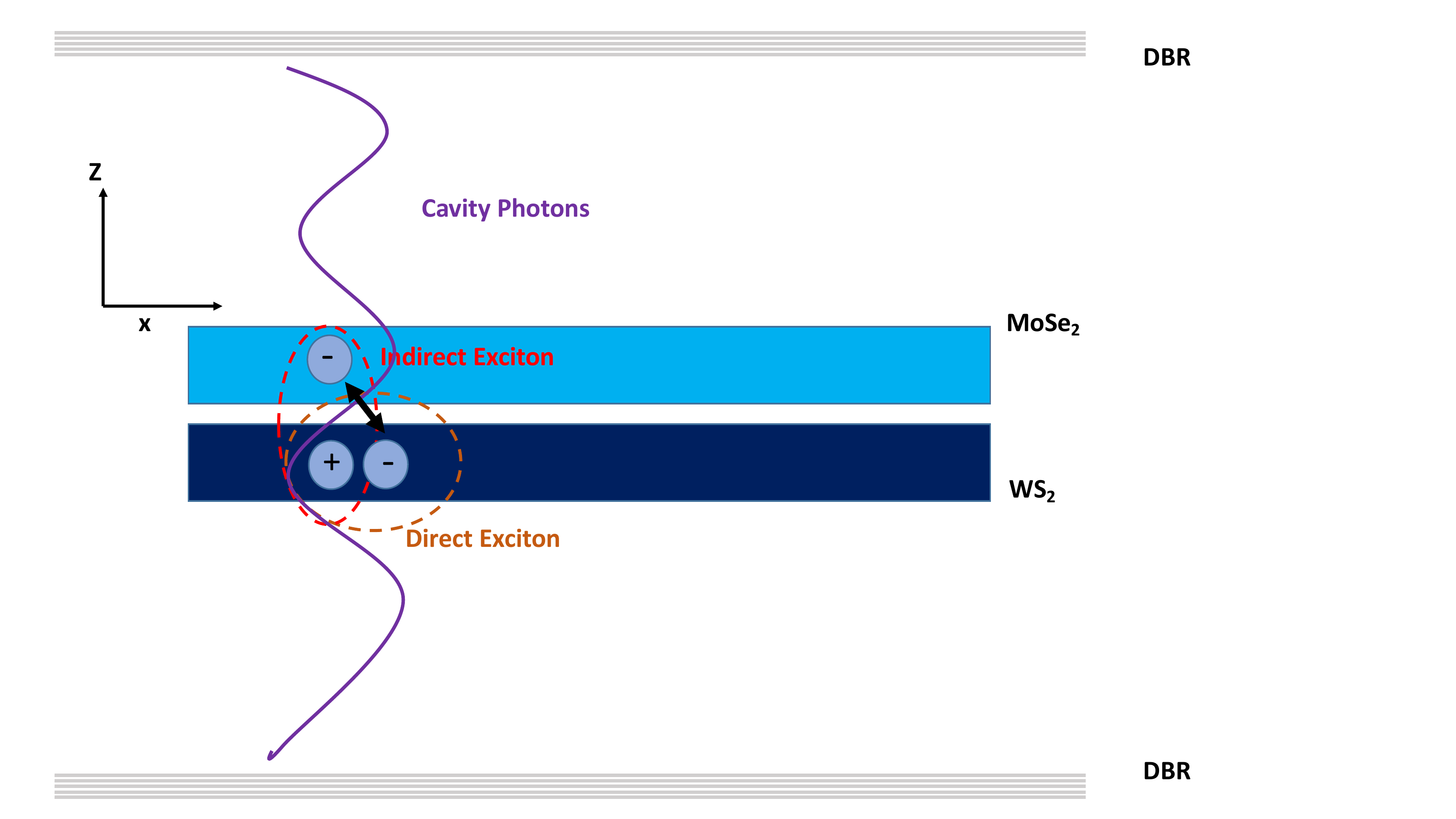}
\caption{Schematic showing the physical process for generating dipolaritons. Spatially separated electrons and electron holes of type II-band alignments between the monolayers of the TMD are photoexcited generating indirect excitons and direct excitons. Coupling of direct excitons with the cavity photons generate a three-way superposition of direct exciton, indirect excitons, and cavity photons. The application of an external voltage provides for a driving force on the dipolaritons which are guided by the Y-shaped channel.  The thick arrow between the indirect exciton and the direct exciton indicates the ability of the electron to tunnel between layers with probability given by the Hopfield coeffcient $|Y|^2$, as described in (8). The dotted lines around the direct exciton and indirect exciton represent bounding by the Coulomb charge. }
\label{fig:pattern}
\end{center}
\end{figure}

\section{Dipolariton diffusion in a TMD heterostructure embedded in an optical cavity}
\label{sec:method}

We consider the motion of dipolaritons under the action of an external electric field generated by a voltage applied to the system acting on the charges located in the MoSe$_2$ layer in the MoSe$_2$-WS$_2$ heterostructure as illustrated in Fig. 2.
The  direct excitons are created by laser radiation in the excitation spot
of characteristic size $\sim 10-20$ $\mu$m at the stem of the channel, 
as described below. 
The interaction of the polaritons with uncoupled excitons can significantly modify the character of the 
polariton motion if they  are located in the same spatial domain \cite{Fernandez:13}.
However, from the simulations below it follows that in our case the dipolaritons travel in the channel 
over the distances
 $\sim 100-500$ $\mu$m  under the action of the electric driving force generated by the application of an external voltage as seen in Fig. 2.
 This distance is much larger than the size of the direct and indirect excitons. Thus, 
the dipolaritons propagate in the area where the uncoupled exciton density is negligible 
and the effect of the exciton-dipolariton interaction can be disregarded. Therefore, in the simulations we omit the exciton-dipolariton coupling in the channel and only consider
 the dipolariton dynamics.

  In this work we follow the approach of modeling the dynamics of the dipolariton gas using the semi-classical stochastic differential equation for the polariton wave packet \cite{Kolmakov:33}: \begin{equation}  
 {\dot\textbf{r}}(t)= \eta_{\text{dip}} \textbf{F}(\textbf{r}(t),t))dt + \sqrt{2D}d\textbf{W}(t), 
 \end{equation} 
 where \textbf{F}(\textbf{r},t) is the external force acting on the dipolaritons, $\eta_{\rm dip}$  is the dipolariton mobility, $D$ is the diffusivity, and $d\textbf{W}(t)$ the differential of a Weiner process \cite{Kolmakov:33}.  The force acting on polaritons is set to $\textbf{F} = -\nabla{U_{\rm eff}{(r)}}$. The laser light spot polariton source is modeled by the addition of particles for each time step $\delta t $ with a Gaussian probability distribution.  The lower dipolariton mass depends on the effective cavity photon mass 
$m_{\rm ph}=\sqrt{\epsilon} \pi \hbar / c L_C$ and the exciton
mass $m_{\rm ex}$=$m_{\rm e}$+$m_{\rm h}$ as follows \cite{Byrnes:14}.
\begin{equation}
{1 \over m}= {|C|^2 \over m_{\rm ph}} + {|X|^2+|Y|^2 \over m_{\rm ex}}.
\end{equation}

Here, $\epsilon$ is the dielectric constant of the host material,
$c$ is the speed of light, $L_C$ is the cavity length,
$m_{\rm e(h)}$ is the electron (hole) mass,  where we take $m_{\rm ex}$=0.7${m}$ with $m_{\rm ex}$ as the free electron mass,  and $m_{\rm ph}$ is the photon mass. The Hopfield coefficients $C$, $X$, and $Y$ for the 
photon, direct and indirect excitons in the dipolariton wave function depend on dipolariton momentum and on 
the detuning in the system. In the case
of zero detuning for both the photons and indirect excitons and low momentum, 
the typical values are $|X|^2=1/2$, $|C|^2=|Y|^2=1/4$ \cite{Byrnes:14}. The dipolariton momentum relaxation time $\tau_{\rm dip}$
depends on the photon $\tau_{\rm ph}$, direct exciton $\tau_{\rm DX}$, and indirect exciton  $\tau_{\rm IX}$ 
lifetimes as
\begin{equation}
{1 \over \tau}_{\rm dip}   = {|C|^2 \over \tau_{\rm ph} } + {|X|^2 \over \tau_{\rm DX}}
+ {|Y|^2 \over \tau_{\rm IX}}.
\end{equation}

In the simulations, the momentum relaxation times for the direct and indirect excitons and cavity photons
were taken as $\tau_{\rm DX} = 4$ ps,  $\tau_{\rm IX} = 80$ ps and $\tau_{\rm ph} = 100$ ps,
respectively \cite{Ceballos:7,Nelsen:13}. The dipolariton diffusivity is calculated as \cite{Kolmakov:33}
 \begin{equation}
D = \frac{m_{\rm ex}}{m}{|X^{-4}|}D_{\rm ex},
 \end{equation}
 where $D_{\rm ex}$  is the exciton diffusion coefficient, with $ X = 1/\sqrt{2} $ as we consider zero detuning between the excitonic and photonic resonances. 
 The dipolariton mobility is calculated as \cite{Kolmakov:33}
 \begin{equation}
 \eta _{\rm dip} = \frac{\tau_{\rm dip}}{m},
 \end{equation}
 for $\tau_{\rm dip}$ where $\tau_{\rm dip}$ is the momentum relaxation time of the dipolaritons.  

 \begin{figure}[h]
\begin{center}
\includegraphics[height=7cm, width=12cm]{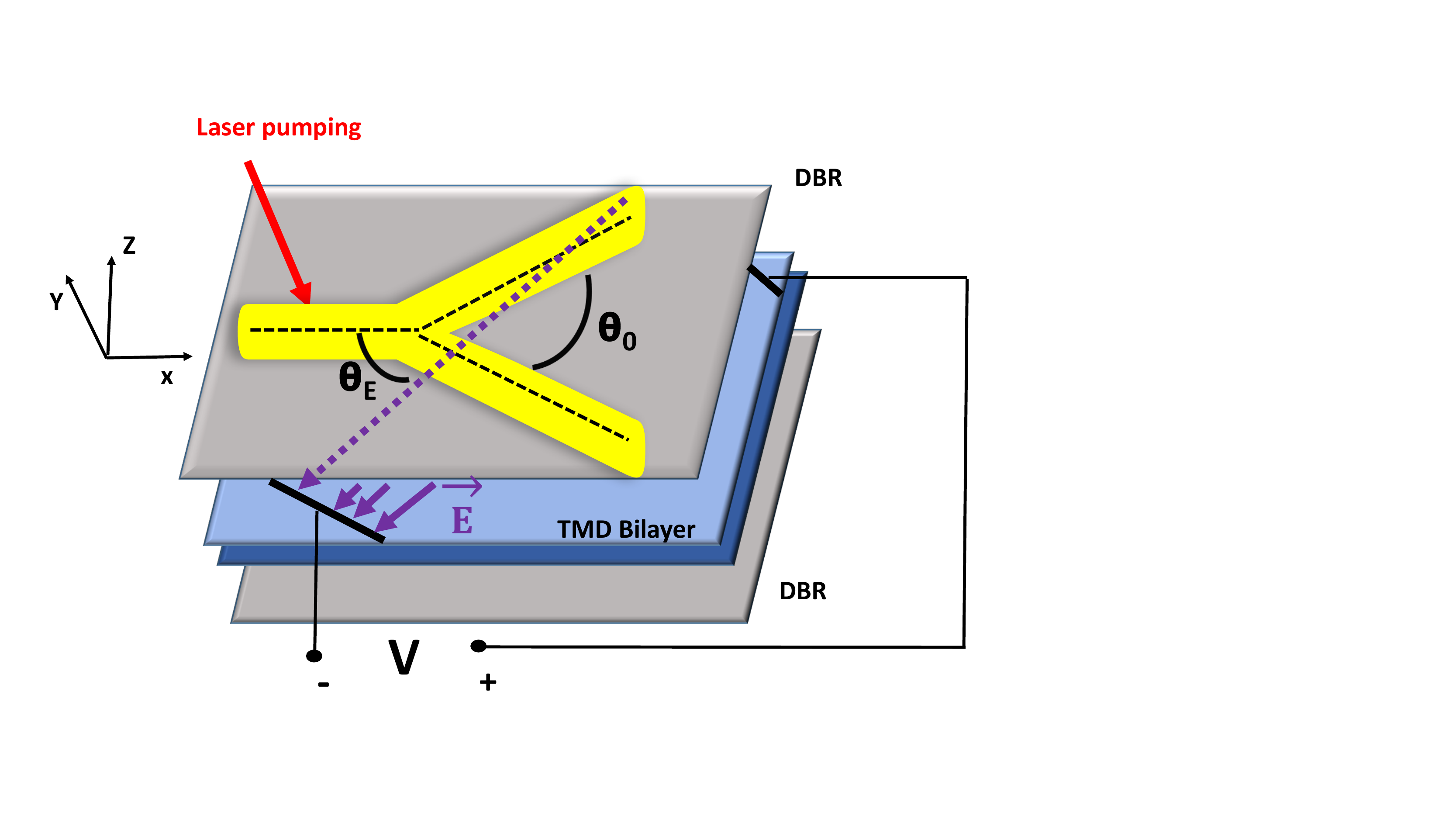}
\caption{Schematic of a TMD heterogenous bilayer embedded inside an optical microcavity with a Y-shaped channel guiding the dipolaritons. The opening angle of the channel is $\theta_0=30$\ and the direction of the field $\bm{E}$ generated by an external voltage applied to the bilayer is defined by an angle $\theta_E$ between the field vector and the direction of the stem of the Y-shaped channel. The direction of the electric driving force applied to the dipolaritons is opposite to the direction of the electric field. Distributed Bragg Reflectors (DBR) are placed between the TMD bilayer and laser pumping is applied to generate dipolaritons, at the beginning of the stem of the channel at which point they propagate along the $x$-axis  towards the junction under the action of the electric field $\bm{E}$.}
\label{fig:pattern}
\end{center}
\end{figure}

The effective potential  $U_{\rm eff}(\bm{r}) $ captures the effects of  
 patterning of the microcavity and of the coordinate dependence of the electrochemical potential 
 of the dipolaritons, 
 \begin{equation}
 U_{\rm eff}(\bm{r}) =  U_{\rm conf} (\bm{r}) + \mu_{\rm e-chem}(\bm{r}).
 \end{equation}
The confining potential due to patterning, $U_{\rm conf}(\bm{r})$ is shown in Fig. 2.
In the case where  the drive voltage is applied across  
the electron-carrying quantum well, the electrochemical potential in the system is 
\begin{equation}
\mu_{\rm e-chem} (\bm{r})= \mu_0 + e |Y|^2 \phi (\bm{r}), \label{eq:echem}
\end{equation}
where $\mu_0$ is the   chemical potential of dipolaritons.
In what follows we consider the case where no voltage is applied across the 
 layer, which carries the holes.
The factor $|Y|^2$ in Eq.\ (\ref{eq:echem}) is  the probability for the electron to be located in the
electron-carrying TMD layer. The effective drive force acting  on the 
dipolaritons  is
\begin{equation}
\bm{F}= - \nabla \mu (\bm{r}) =  e  |Y|^2 \bm{E}, \label{eq:force}
\end{equation}
where the electric field in the system  is $\bm{E} = - \nabla \phi (\bm{r})$. 
In this paper, we consider the simplest case 
where the electric field $\bm{E}$ is uniform. 
The source of the dipolaritons $P(\bm{r})$  was taken in the form of a Gaussian function 
centered  at the base of the stem (see Fig.\ \ref{fig:pattern}) with the full width at half maximum
(FWHM) of 16.7 $\mu m$

\begin{figure}
\includegraphics[height=5cm, width =9cm]{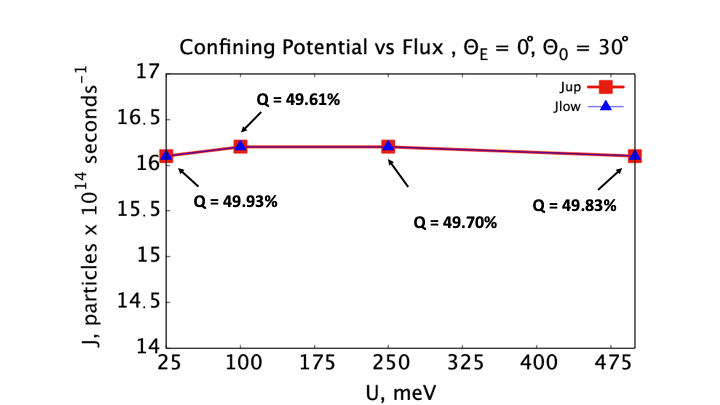}

\caption{Confining potential $U$ versus flux $J$ for the case of the opening angle of the channel, $\theta_{0}$= 30{$^{\rm o}$} and the electric field angle, $\theta_{E}$ = 0{$^{\rm o}$}. The flux through the upper branch of the Y-shaped channel is labeled as $"J_{\rm up}"$ and the flux through the lower branch of the Y-shaped channel is labeled as $"J_{\rm low}"$. The performance $Q$ is indicated by the arrows pointing to the overlapping flux points.}
\end{figure}

\begin{figure}[htp]
\centering
\includegraphics[width=.5\textwidth]{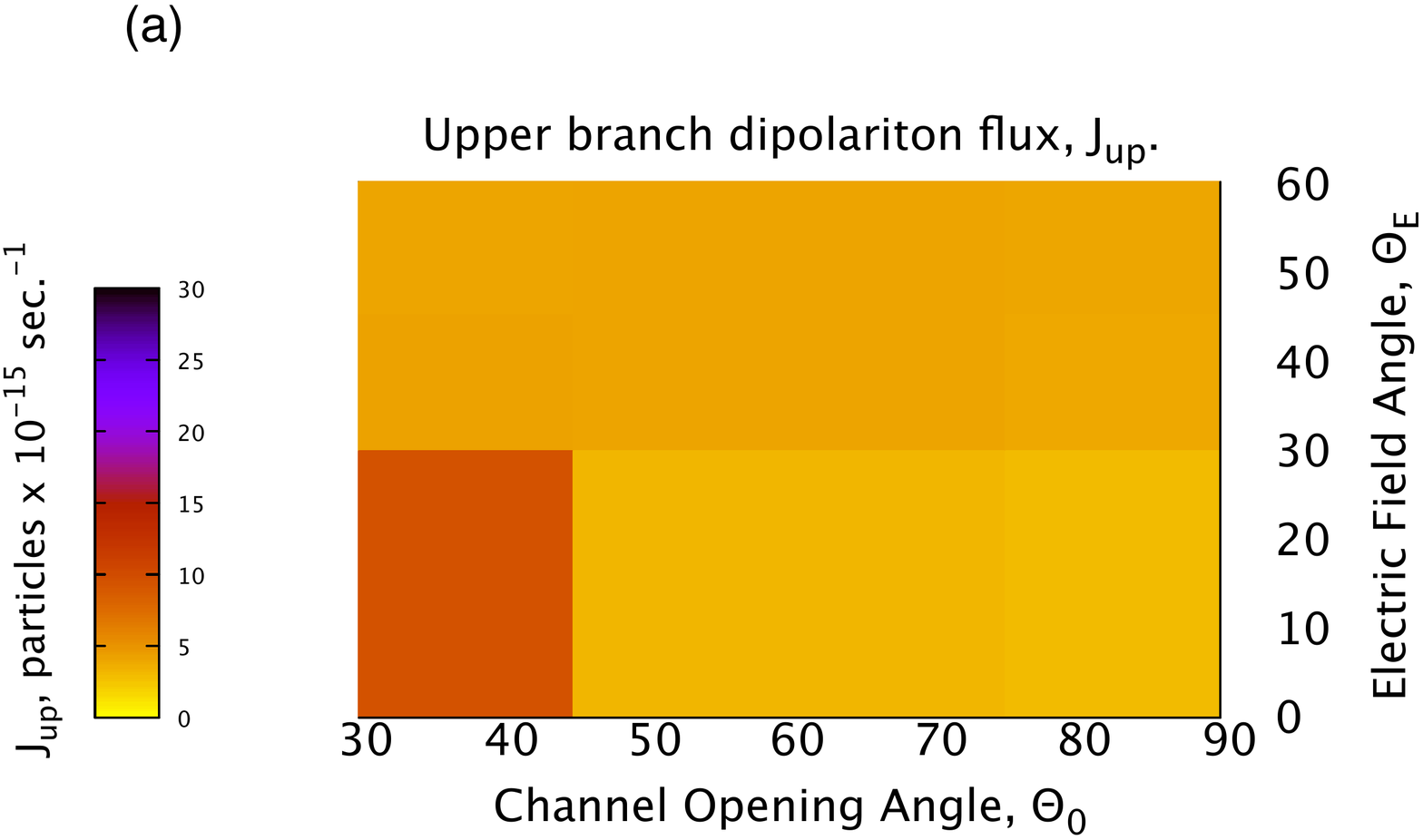}
\includegraphics[width=.5\textwidth]{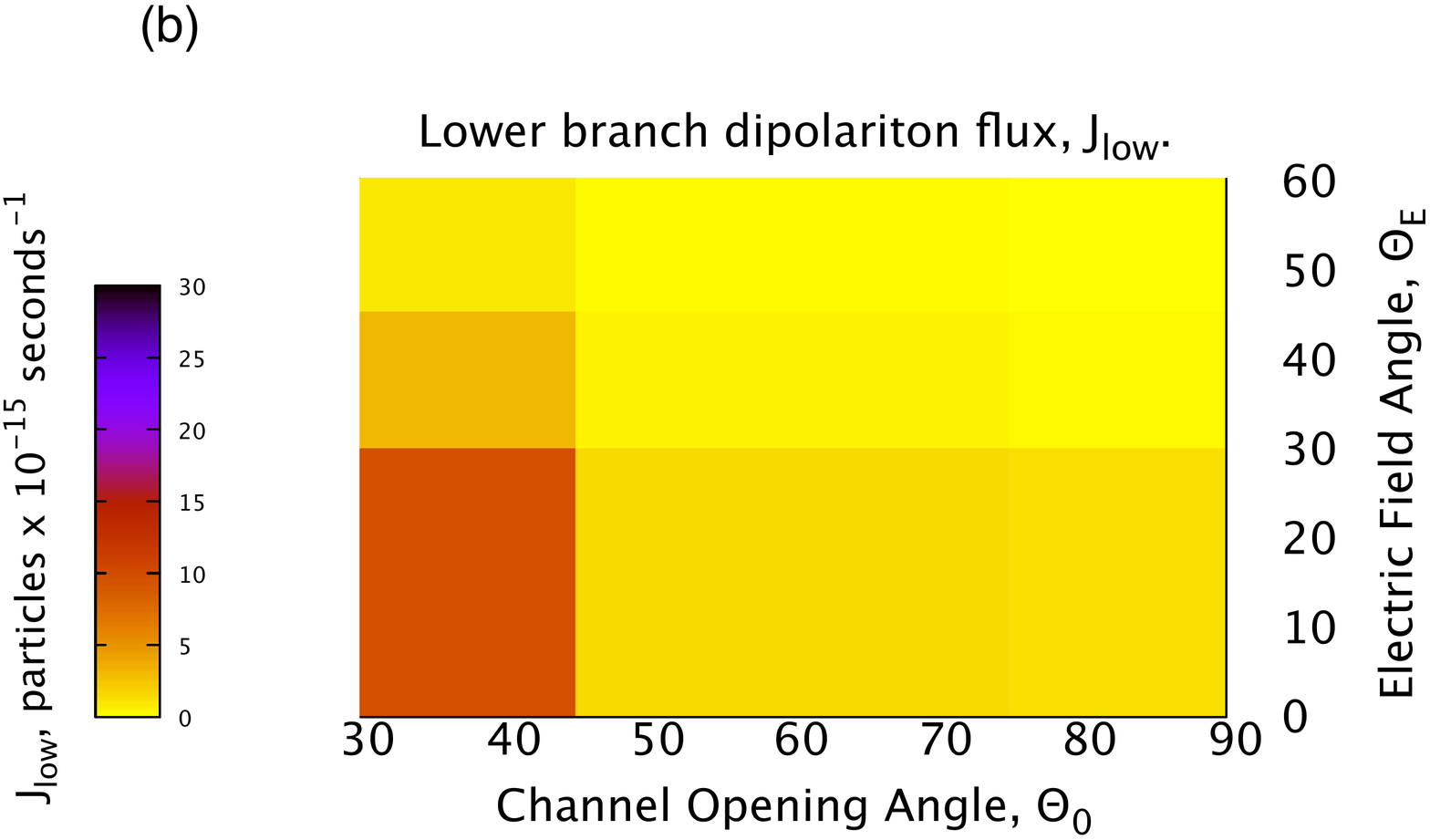}
\includegraphics[width=.5\textwidth]{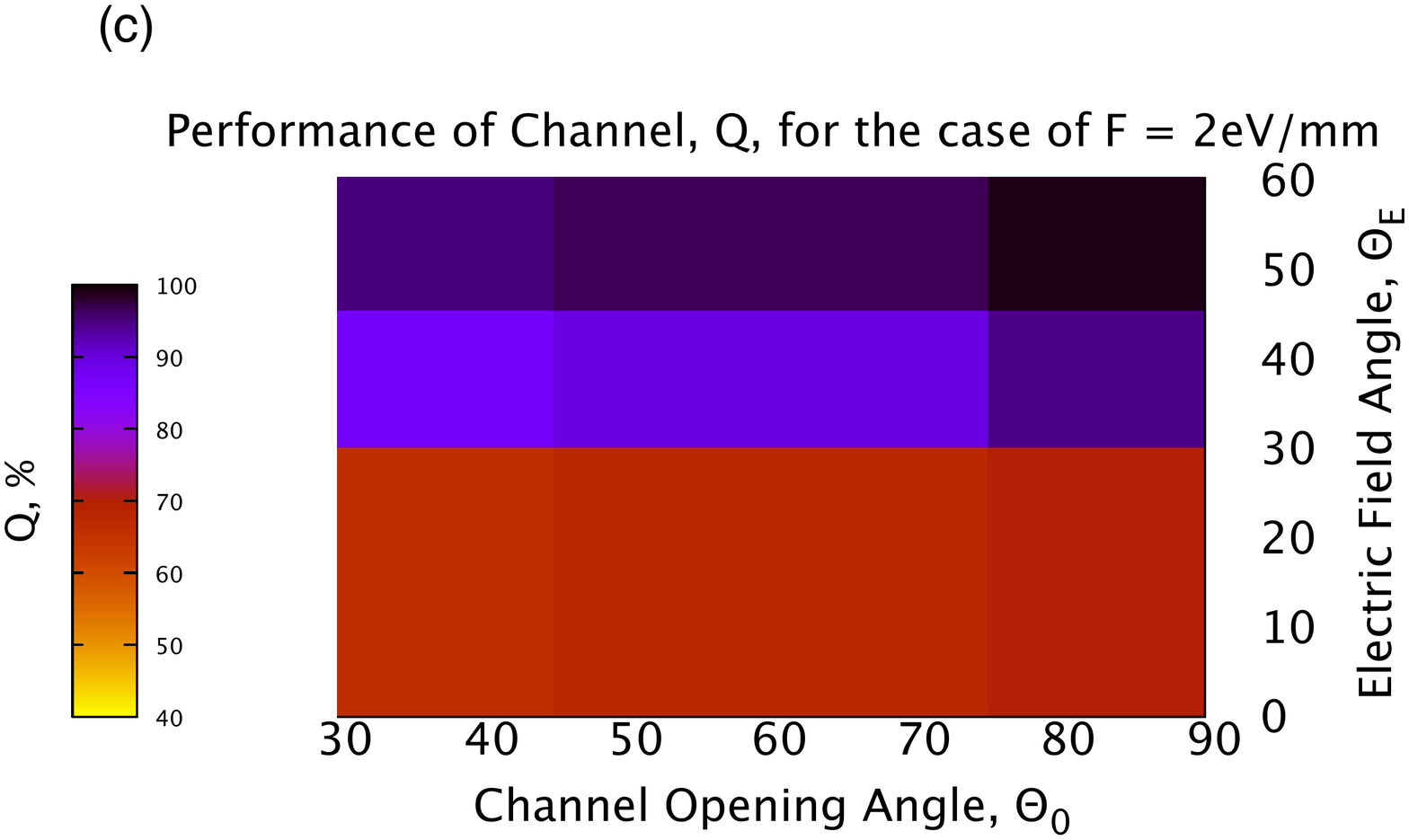}
\caption{ (a) the upper dipolariton flux, (b) lower dipolariton flux in the Y-channel as functions of the channel opening angle and electric field angle with the electric driving force $F$ set to 2eV/mm.  (c) The performance of the channel as a function of the opening angle and electric field angle with the driving force set to $F$ = 2eV/mm. The performance in (c) as defined in (11) is calculated using the upper and lower dipolariton flux plots.}
\end{figure}

Furthermore, we consider zero detunings where the
 cavity
photons and the excitons are in resonance at $\bm{k}=0$.

\begin{table}[tb]
\centering
\caption{Simulation parameters for a cavity with embedded MoSe$_2$-WS$_2$  bilayer}
\begin{tabular}{|c|c|c|} 
\hline 
\underline{Quantity name} & \underline{Value} & \underline{Variable} \\

Exciton mass & $m_{\rm ex}$ & $0.70 m_{\rm 0}$ \\
Photon mass & $m_{\rm ph}$ & $1.1234\times 10^{-5} m_{\rm 0}$ \\
Dipolariton mass & $m$ & $2.4 \times 10^{-5} m_{\rm 0}$ \\
Dipolariton lifetime  & $\tau_{\rm dip} $ & $15.64 \times 10^{-12}$ s \vspace{-0.1cm}\\
\hspace{0.15cm}  & & \\
Indirect exciton lifetime  & $\tau_{\rm IX} $ & $80 \times 10^{-12}$ s \vspace{-0.1cm}\\
\hspace{0.15cm}  & & \\
Direct exciton lifetime  & $\tau_{\rm DX}$ & $4.0 \times 10^{-12}$ s \vspace{-0.1cm}\\
\hspace{0.15cm}  & & \\
Cavity Photon lifetime  & $\tau_{\rm ph} $ & $100\times 10^{-12}$ s \vspace{-0.1cm}\\
\hspace{0.15cm}  & & \\
Exciton diffusion   & $D_{\rm ex}$ &  $14$ cm$^2$/s \vspace{-0.1cm}\\
\hspace{0.15cm} coefficient & & \\
Dimensionless dipolariton    & $D \times dt/dx^{2}$  &  $271.7$  \vspace{-0.1cm}\\
\hspace{0.15cm} diffusion coefficient & & \\

Dimensionless dipolariton & $\eta_{\rm dip} \times eVdt/dx$ &  $0.0015$  \vspace{-0.1cm}\\
 \hspace{0.15cm} mobility   & & \\
Dielectric constant & $\epsilon$ & 4 \\
Exciton Energy & $E_{\rm ex}$ & 1.58eV \\
Confining potential  & $U$ & $25-500 meV$\\
Numerical unit of  & $dx$ & 0.15 $\mu$m \vspace{-0.1cm}\\
\hspace{0.15cm}length & & \\
Numerical time step & $dt$ & 9.63 fs \\
\hline
\end{tabular}
\label{tab:params}
\end{table}

To characterize the flow of the dipolariton condensate in the channel,
we calculated the total dipolariton flux through the upper (lower) branches of the Y-channel junction
\begin{equation}
J = \int da j_{||}, \label{eq:jn}
\end{equation}
where the integration is performed  along the cross-section of the branches, $da$ is the area being integrated over and   
$j_{||} = \bm{j} \cdot \bm{\nu}$ is the 
component of the dipolariton flux along the channel. The angles $\theta_{0}$ and $\theta_{E}$ are only included in our numerical modeling of the Y-channel structure that the condensate flows in and does not find direct applicability in Eq.(9).

 The flux in our simulation is found by simple averaging of particles. We adopt the notation of $ \Delta J dt = \pm 1 $ and define $J_{\rm up}{ dt}$ = ${+1}$ and $J_{\rm low}{ dt} = {-1}$, where $\pm 1$ is taken in numerical units, with  1 numerical unit  =  $9.6\times 10^{15}$ dipolariton particles per second. When the system comes to a steady state, particles found at  $x$ $>$  $x_{\rm j}$, $y$ $>$ $y_{\rm j}$ are counted as $J_{\rm up}$,  where $x_{\rm j}$ = 450$\mu$m,  $y_{\rm j}$ = 225$\mu$m. The counting locations for $x_{\rm j}$ and $y_{\rm j}$ were selected as to ensure the particles would be at a location past the junction of the Y-channel to accurately record them as being either through the upper or lower branch of the Y-channel, rather than at some location along the stem of Y-channel. Particles found at  $x < x_{j}$, $y > y_{j}$,   are counted as $J_{\rm up}$. Particles found at  $x > x_{\rm j}$, $y < y_{\rm j}$ are counted as $J_{\rm low}$ and particles found at  $x < x_{\rm j}$, $y < y_{\rm j}$ are counted as $J_{\rm low}$. More compactly, this counting scheme can be expressed as 

\begin{equation}
\Delta Jdt= \text{sgn}(x-x_{\rm j})\text{sgn}(y-y_{\rm j}) \label{eq:s}.
\end{equation}

\begin{figure}[htp]
\centering
\includegraphics[width=.5\textwidth]{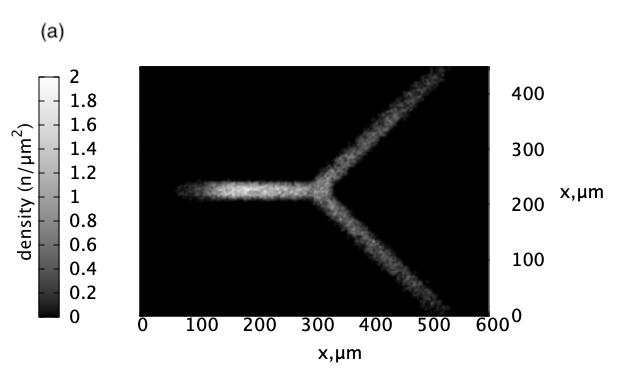}
\includegraphics[width=.5\textwidth]{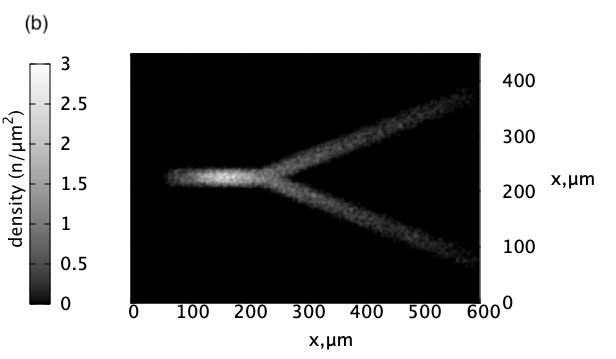}
\caption{Steady-state dipolariton condensate flow in at the Y-shaped channel in a MoSe$_2$-WS$_2$  based microcavity for (a) an opening angle of $\theta_{0}$ = 90{$^{\rm o}$}, driving force $F$ = 1.0eV/mm, electric field angle $\theta_{0}$ = 0{$^{\rm o}$} and (b) an opening angle of $\theta_{0}$ = 45{$^{\rm o}$}, driving force $F$ = 0.5eV/mm, electric field angle $\theta_{0}$ = 90{$^{\rm o}$}. The gradient bars on the left of the plots labeled with units  ${n}/{{\mu m}^2}$,  the number of particles per micron-squared, display the dipolariton condensate density.}
\end{figure}

 To characterize the redistribution of the dipolariton flux in the channel 
 in response to  $F$, we calculated 
 the fraction of dipolaritons propagating through the upper branches, or what we define as the performance,  as 
 \begin{equation}
 Q={J_{\rm up} \over {J_{\rm up} + J_{\rm low}}}  \times 100\%. \label{eq:q}
 \end{equation}
 
 We define $Q$ in such a manner in order to find the percentage of dipolariton flux going through the upper branch of the channel relative to the overall dipolariton flux in the Y-shaped channel. This enables us to quantify to what extent we can re-route the total dipolariton flux in the channel through the upper branch of the Y-shaped channel.
 
In the simulations, we set the maximum depth of confining potential $U_{\rm conf}(\bm{r})$ equal to 250 meV.
To study the effect of the confining potential depth on the dipolariton flow we varied
its value from 25 meV to 500 meV. We found statistically insignificant differences in performance $Q$, as defined in Eq.(11), and dipolariton fluxes, $J$ as defined in Eq.(10)  for this range of confining potential as shown in Fig 3. Thus, we can claim that our results very weakly depend on the confining potential. The simulation parameters are shown in Table~1.

\begin{figure}[t]
\includegraphics[width=10cm, height=7cm]{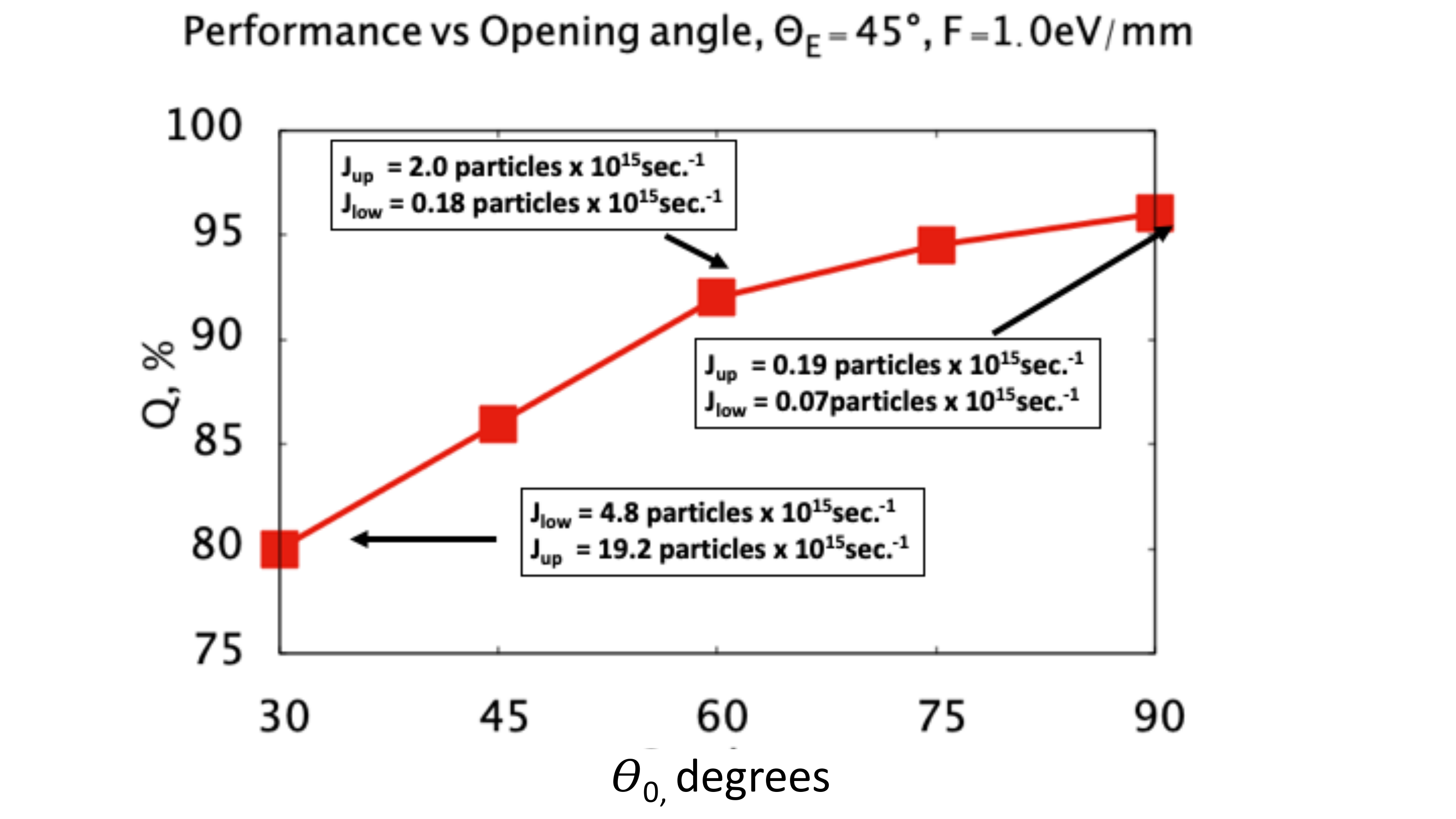}
\caption{Performance $Q$ versus opening angle, $\theta_{0}$, for the case $\theta_{E}$ = 45{$^{\rm o}$}  and $F$ = 1.0eV/\text{mm}. Arrows point to the fluxes, $J_{\rm up}$ and $J_{\rm low}$, for corresponding $Q$ values.}
\end{figure}

\begin{figure}[t]
\includegraphics[width=9cm, height=11cm]{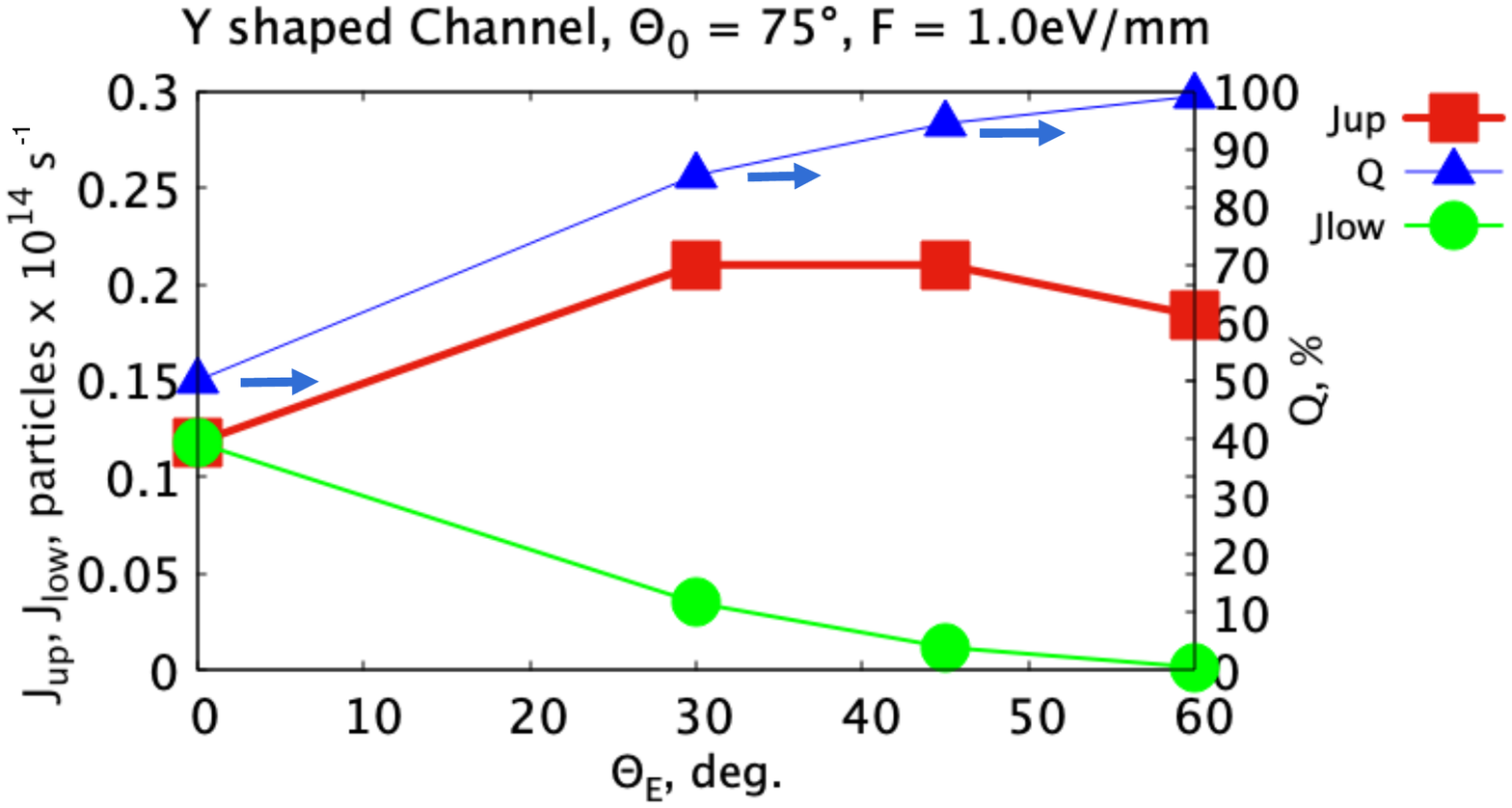}
\caption{Dipolariton flux through the upper and lower branches of the Y-shaped channel,  $J_{\rm up}$ and $J_{\rm low}$, and the $Q$ factor as functions of the direction of the in-plane electric field, $\theta_{E}$, for the case $\theta_{0}$ = 75{$^{\rm o}$}  and F = 1.0eV/mm. The blue arrows point to the $y_{2}$ axis indicating flux for the value.} 
\label{fig:gaaso120}
\end{figure}

\section{Optimizing the performance of the Y-channel to effectively reroute dipolaritons}
In order to determine the optimal condition for directing dipolaritons in the Y-shaped channel, we varied $\theta_{0}$, the opening angle of the channel, from $\theta_{0}$ = 30{$^{\rm o}$}  to $\theta_{0}$ = 90{$^{\rm o}$}. The cases of  $\theta_{0}$ $ < $  30{$^{\rm o}$} were not investigated as the $\theta_{0}$ = 30{$^{\rm o}$} was chosen to be the lowest opening angle that still maintains our channel classification as one possessing a Y-shaped channel. The plots in Fig. 5 illustrate the condensate density for the Y-shaped channel and serve as a visual representation of the condensate when varying system parameters in our optical microcavity.  The behavior of the channel can be summarized in Fig.4(a)-(c) where we can observe the dipolariton fluxes and the channel performance as functions of channel opening angle and electric field angle. We can see in Fig. 4(c) that the performance $Q$ is optimized when the electric field angle and opening angle are maximized. Further, we can see that $Q$ has values greater than 90$\%$ when the electric field angle is at 60{$^{\rm o}$}  irrespective of the field angle when the driving force is greater than 2eV/mm. It is interesting to note, however, that both upper and lower dipolariton flux is maximized at channel opening angles and electric field angles of 60{$^{\rm o}$} . This can be understood by noting that an increased opening angle requires the dipolaritons to travel farther from the initial excitation spot in order to pass through the branches of the channel with an increased opening angle. In what follows we will investigate and cite results for particular case studied; that is, cases where the electric field angle $\theta_{\rm E}$, the channel opening angle $\theta_{0}$, and electric driving force $F$, are varied. In particular, we numerically calculated the  performance $Q$ as a function of $\theta_{0}$ for the case of $\theta_{E}$ = 45{$^{\rm o}$} and $F$ = 1.0eV/mm. As illustrated in Fig. 6, we can see that $Q$ is maximized for the case of $\theta_{0}$ = 90{$^{\rm o}$ where $Q$ reaches a value of $\approx$ 96$\%.$ Furthermore, we see that the lowest value of the opening angle, $\theta_{0}$ = 30{$^{\rm o}$} , provides for a minima for this case where the performance reaches $Q$ $\approx$ 80$\%$. Thus, we can claim that $Q$, the performance, monotonically increases with the opening angle of the channel for the range we investigated of $\theta_{0}$ = 30{$^{\rm o}$}  to 90{$^{\rm o}$}  where we can observe that performance can be improved by up to 16\%. It is noteworthy, however, to observe that the flux of particles through the upper and lower branch, $J_{\rm up}$ and $J_{\rm low}$, decreases with increase of $\theta_{0}$ as seen in Fig. 6. We can see that the flux of particle for the case of $\theta_{0}$  = 30{$^{\rm o}$}  is substantially higher than the flux of particles for the other opening angles. In particular, from Fig. 6 we can see that for $\theta_{0}$  = 30{$^{\rm o}$}   $J_{\rm up}$ and $J_{\rm low}$ reach values of 19.2 $\times$ $10^{15}$ $\text{s}^{-1}$ particles and 4.8 $\times$ $10^{15}$ $\text{s}^{-1}$particles, respectively, compared to $\theta_{0}$  = 90{$^{\rm o}$} where $J_{\rm up}$ and $J_{\rm low}$ reach values of $0.19\times 10^{15} \text{s}^{-1} $particles and $0.07\times 10^{15} \text{s}^{-1}$ particles, respectively. Thus, an increase in the opening angle of the channel lowers the flux of dipolaritons in the channel while increasing the performance.

 In order to test efficacy of increasing $\theta_{0}$ on $Q$ for different parameters, we numerically computed $Q$ as a function of $\theta_{0}$ for the case of $\theta_{E}$ = 0{$^{\rm o}$ and $F = 1.0 $eV/mm$ $. It was found that for a value of $\theta_{\rm E}$ = 0{$^{\rm o}$, the increase in $Q$, is minimal, although still monotonically increasing with $\theta_{E}$. We can observe an increase in performance of  $\Delta Q\approx$1.3$\%$  as $\theta_{0}$ = 30{$^{\rm o}$ is increased to $\theta_{0}$ = 90{$^{\rm o}$. Thus, we can claim that an increase in performance from a wider opening angle of the channel must be accompanied by a non-zero electric field angle in order to appreciate $Q$ significantly.

In order to find an optimum in $Q$ as a function of the electric field angle, $\theta_{\rm E}$, we numerically calculated the performance $Q$ of the Y-shaped channel as a function of the electric field angle $\theta_{\rm E}$. Fig. 7  shows the fluxes $J_{\rm up}$ and $J_{\rm low}$\ and the performance $Q$ as function of the electric field angle $\theta_{\rm E}$. We can see that both $Q$ and $J_{\rm up}$ are a monotonically increasing functions with $\theta_{E}$ whilst $J_{\rm low}$ is a montonically decreasing function with $\theta_{\rm E}$. In particular, we can claim an increase in performance of  $\approx$50$\%$ as we increase the angle of the electric field from  $\theta_{\rm E}$ = $\ang{0}$ to $\theta_{\rm E} = \ang{60}$. Fig. 7 shows that the flux through the upper branch, $J_{\rm up}$ is increased by $\approx$ $0.63$ $ \times 10^{15}$ $\text{s}^{-1}$ particles as we increase the angle of the electric field from  $\theta_{\rm E}$ = $\ang{0}$ to $\theta_{E} = \ang{60}$. Inspection of Fig. 7 reveals that the most performance per increase in electric field angle occurs in the range of $\theta_{\rm E} = \ang{0}$ to $\theta_{\rm E}= \ang{30}$ as $Q$ is appreciated by  $\approx$31$\%$. The increase of  $\theta_{E}$ from $\ang{30}$ to $\ang{45}$ increases performance by $\approx 9$\% and the increase of  $\theta_{\rm E}$  from \ang{45} to \ang{60} increases performance by $\approx$5$\%$. Thus, for practical matters, the most substantial increase in performance occurs in the range $\theta_{\rm E} = \ang{0}$ to $\theta_{\rm E} = \ang{30}$ with an optimum in performance for $\theta_{\rm E} = \ang{60}$.
Thus, we can claim that an increase of the electric field angle appreciates the performance to a higher extent than increasing the opening angle and increases the flux of dipolaritons in the channel.

\begin{figure}[t]
\includegraphics[width=10cm, height=7cm]{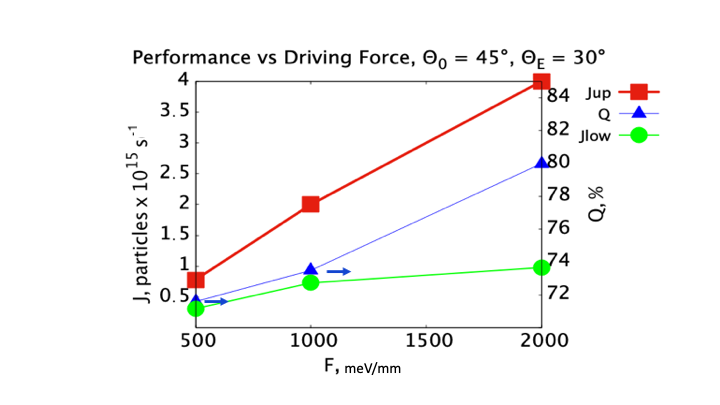}
\caption{Dipolariton flux through the upper and lower branches of the Y-shaped channel,  $J_{up}$ and $J_{low}$, and the $Q$ factor as functions of the direction of the driving force, F, for the case $\theta_{0}$ = \ang{45} and $\theta_{E}$ = \ang{30} . The blue arrows point to the $y_{2}$ axis indicating flux for the value} 
\label{fig:gaaso120}
\end{figure}

In order to find an optimum in $Q$ as a function of the driving force $F$ on dipolaritons, we numerically calculated the performance of the Y-shaped channel as a function of the driving force on dipolaritons. In Fig. 8 we can can see the performance as a function of the driving force. In particular, we see that Q, $J_{\rm up}$ and $J_{\rm low}$ are monotonically increasing  with $F$. In particular, we can observe an improvement of performance of  $\approx$ 8.5$\%$ for an increase in driving force from 500meV/\text{mm} to 2000meV/\text{mm}. We can also report an increase in the flux of particles, $J_{\rm up}$ and $J_{\rm low}$, of  $\approx$ 3.3$\times$ $10^{15}$ $\text{s}^{-1}$ particles and 0.7$\times$ $10^{15}$  $\text{s}^{-1}$ particles, respectively. This dependence has also been observed for other opening angles and electric field angles.  Thus, we can claim that an increase in the driving force on the channel can increase both the flux of dipolaritons and the performance of the channel.

Based on our variation of channel parameters, we were able to find optimal conditions for performance. We report these optimal parameters in Fig. 9 where we found our optimum conditions as $\theta_{E}$ = 60{$^{\rm o}$, $\theta_{0}$ = 90{$^{\rm o}$, and $F$ = 2.0eV/mm. We can report that $Q$ reaches a maxima of $\approx$ 100$\%$ when $\theta_{E}$ = 60{$^{\rm o}$, whilst the minima of this case is in the usual vicinity of $\approx$ 50$\%$. Furthermore, it is noteworthy that the performance reaches $\approx$ 92$\%$ for the electric field angle of $\theta_{E}$ = 30{$^{\rm o}$, which is larger than the cases in Fig. 9, where $\theta_{E}$ is held fixed at  $\theta_{E}$ = 30{$^{\rm o}$. 

\vspace{0.8cm}

\section{Conclusions}\vspace{0.2cm}
By considering dipolariton propagation in a Y-shaped TMD channel embedded in an planar optical microcavity, we demonstrated that the dipolariton flow  can be efficiently re-routed  by applying the electric field driving force $\sim 2$ eV/mm at angle $ \theta_{E}$ = \ang{60} and an opening angle of  $\theta_{0}$ = \ang{90}. As TMDs have been shown to function at room temperature  \cite{calman:11}, the dipolariton switch we have investigated will likewise be able to operate at room temperature. There is an optimum in the angle of the electric field direction, $\theta_{E}$, for the Y-shaped channel for which the value of $Q$ is maximized.  Thus, when the electric force acting on dipolaritons is 2eV/mm with $\theta_{E}$ = \ang{60}, about 100\% of dipolaritons can be switched in the desired direction in the channel for all opening angles of the channel.

The value of $Q$ monotonically increases with an increase of $\theta_{E}$ for most cases investigated in the Y-channel; in particular, this monotonicity is only broken for $\theta_{0}$ = \ang{60} where the conditions in the channel are  $F = 2$eV/mm,  $\theta_{E}$ = \ang{60}, $F = 2$ eV/mm, $\theta_{E}$ = \ang{45}, and $F = 1$ eV/mm, $\theta_{E}$ = \ang{60}. This can attributed to the stochastic and random nature of the system in question. Outside the optimal condition parameter range, the efficiency $Q$ and dipolaritons cannot be efficiently re-routed in the channel. Our consideration opens a route to the design of efficient room-temperature optoelectronic applications, including optical routers and switches, based on dipolaritons in TMD microcavities.

\begin{figure}[t]
\hspace{0.5cm}\includegraphics[width=9cm, height=8cm]{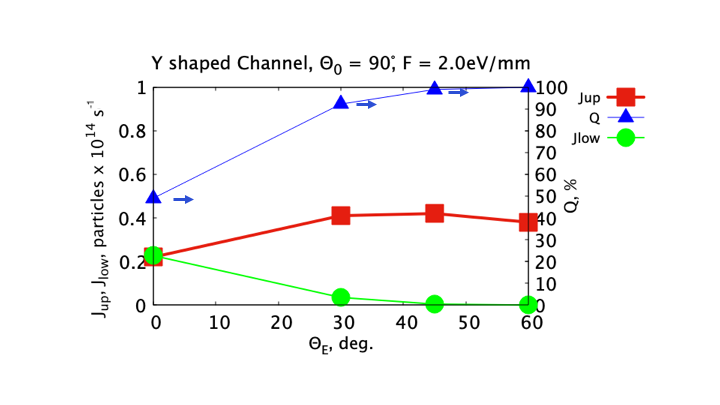}
\caption{Dipolariton flux through the upper and lower branches of the Y-shaped channel,  $J_{up}$ and $J_{low}$, and the $Q$ factor as functions of the electric field angle, $\theta_{E}$,   for the case $\theta_{0}$ = \ang{90} and F = 2.0eV/mm . The blue arrows point to the $y_{2}$ axis indicating flux for the value} 
\end{figure}

\section{Acknowledgments}
This work was supported in part by the Department of Defense under the grant No. W911NF1810433. The authors are grateful to The Center for Theoretical Physics New York City College of Technology of The City University of New York for providing computational resources. The authors are also grateful to R. Ya. Kezerashvili and O. L. Berman for fruitful discussions.

\bibliographystyle{unsrt}
\bibliography{Arxiv_dipolaritonswitch.bib}

\end{document}